# Agricultural Roots of Social Conflict in Southeast Asia

Justin Hastings[*]    David Ubilava[†]

**Abstract**

We examine whether harvest-time transitory shifts in employment and income lead to changes in political violence and social unrest in rice-producing croplands of Southeast Asia. Using monthly data from 2010 to 2023 on over 86,000 incidents covering 376 one-degree cells across eight Southeast Asian countries, we estimate a general increase in political violence and a decrease in social unrest in croplands with rice production during the harvest season relative to the rest of the crop year. In a finding that is least sensitive to alternative model specifications and data subsetting, we estimate a nine percent increase in violence against civilians in locations with considerable rice production compared to other parts of the region during the harvest season, relative to the rest of the year. We show that the harvest-time changes in conflict are most evident in rural cells with rainfed agriculture. Using location-specific annual variation in growing season rainfall, we then show that the harvest-time increase in violence against civilians occurs in presumably good harvest years, whereas increase in battles between actors of political violence follows growing seasons with scarce rainfall. The harvest-time decrease in social unrest, protests in particular, occurs after presumably bad harvest years. These findings contribute to research on the agroclimatic and economic roots of conflict and offer insights to policymakers by suggesting the spatiotemporal concentration of conflict as well as diverging effects by forms of conflict at harvest time in the rice-producing regions of Southeast Asia.

[*] Discipline of Government and International Relations, University of Sydney.
[†] School of Economics, University of Sydney. Corresponding author: david.ubilava@sydney.edu.au

**Introduction**

In low- and middle-income countries, a small change in people's income may trigger behavioral responses, some of which may be unlawful as well as violent. While it is common for such incidents to occur in cities, where most people live and the state administration resides (e.g., Smith, 2014; Hendrix and Haggard, 2015), social conflict is not just an urban phenomenon. Rural areas, which typically encompass territories with limited state capacity, also experience conflict due to changes in income and employment. Mounting empirical evidence points to a connection between agriculture and conflict (Wischnath and Buhaug, 2014; Koren, 2018; Vestby, 2019; McGuirk and Burke, 2020; Linke and Ruether, 2021), wherein agricultural output is considered a mechanism connecting climate shocks with conflict (e.g., Burke et al., 2009; Hsiang et al., 2013; Dell et al., 2014; von Uexkull et al., 2016; Crost et al., 2018; Koubi, 2019; Buhaug and von Uexkull, 2021; Vesco et al., 2021).

Does conflict intensify at harvest in locations with crop agriculture? The answer to this question is not trivial because of the complex set of motives governing this relationship through multiple channels. There is a burgeoning literature linking agricultural income with political violence and social unrest (e.g., Hendrix and Salehyan, 2012; Koren and Bagozzi, 2017; Linke and Ruether, 2021; Koren and Schon, 2023), often concluding that conflict can lead to food insecurity, and that agricultural shocks are associated with changes in the frequency and intensity of conflict.

To address the research question, we examine over 86,000 incidents across eight Southeast Asian countries during the 2010–2023 period. We differentiate between incidents involving different types of conflict actors: organized armed groups engaged in political violence and unorganized groups engaged in social unrest. In a finding that is least sensitive to alternative model specifications and data subsetting, we estimate a nine percent increase in violence against civilians in locations with considerable rice production compared to other parts of the region during the harvest season, relative to the rest of the year. By focusing on the specific time of the year in areas with a relative surplus of agricultural spoils, and by analyzing a subset of the data involving a specific group of predators whose goal is to achieve a relative advantage through their violent actions, we present suggestive evidence that the harvest season presents rapacious groups with an opportunity to seize or destroy agricultural output. In support of this evidence, we also show that the effect primarily stems from rural areas with rainfed agriculture.

This study contributes to our understanding of the relationship between agriculture, climate, and conflict in several respects. First, the literature on the agricultural roots of conflict is often equivocal (Berman et al., 2011; Koren and Bagozzi, 2017; Koren, 2019; Crost and Felter, 2020; McGuirk and Burke, 2020; Grasse, 2022), in part because agriculture and conflict can be linked through different mechanisms, resulting in different outcomes. By focusing on different types of actors, we suggest that different forms of conflict can have different economic logic. We develop a theory of harvest-time conflict that looks at the pathways by which the harvest can lead to changes in conflict intensity: groups may be acting out of resentment at inequality with others, out of rapacity at the opportunity to destroy or appropriate agricultural surplus, or because of the transitory changes in the opportunity cost of conflict during harvest times. We enrich literature by presenting evidence for the potentially diverging effects that agricultural windfalls can have on different forms of conflict.

Second, we bring a focus on Southeast Asia, a relatively understudied region (Adams et al., 2018) to the literature on the climatic roots of conflict (e.g., Burke et al., 2009; Hsiang et al., 2013; Dell et al., 2014; Crost et al., 2018), and add important nuance to the relationship between the *timing* of climate shocks and conflict. We present empirical support for the effect of growing season precipitation on conflict through its effect on the harvest of the dominant staple cereal crop in the region. We show that growing season deviations in rainfall can have varying harvest-time effects on different forms of conflict that are likely governed by different mechanisms, including those not necessarily mediated through agricultural windfalls.

Third, we contribute to the emerging literature on the seasonality of conflict (Harari and La Ferrara, 2018; Ubilava et al., 2023; Guardado and Pennings, 2024; McGuirk and Nunn, 2024) by showing that conflict seasonality varies by type of conflict and takes place in a larger political context. Much of the literature is focused on the effect of food shortages or food insecurity on conflict, finding that fluctuating food prices can lead to violent conflict and social unrest (Bellemare, 2015; Chen et al 2018), and more food insecurity leads to armed conflict (Koren and Bagozzi 2016). We focus instead on the connection between agricultural *windfalls* and conflict. Using harvest season as the period of intermittent employment and income, we estimate an increase in conflict by organized groups and a reduction in conflict by less organized actors, and thus provide new empirical evidence for existing theories of conflict.

**Southeast Asia as a Case**

We choose Southeast Asia, rather than all of Asia for example, as a suitable region for the analysis for several reasons. First, politically, it is a coherent region for comparison purposes, inasmuch as all Southeast Asian countries (bar Timor-Leste) are members of the Association of Southeast Asian Nations (ASEAN), and as such are dedicated to peaceful resolution of conflicts with each other, mutual development, and closer political, economic, and social cooperation.

Second, Southeast Asia mostly consists of low- and middle-income countries, with a significant portion of their population living at or below the national poverty line (Viana et al., 2022; World Bank, 2022a, 2022b). The region also has large between- and within-country variations in governance and institutional capacity levels, as the Philippines, Myanmar, and Indonesia in particular struggle to control their geographic peripheries.

Third, agriculture is an important sector for employment and income generation across much of the region (World Bank, 2022c). The region's countries also have broadly similar climate, farming practices, and agricultural production patterns (as compared with Asian regions at more northern latitudes). While subsistence farming is prevalent in Southeast Asia, several countries in the region are also net food exporters (which differs from many African countries). Thailand, Vietnam, Myanmar, and Cambodia are among the top 10 rice-exporting countries globally (FAO, 2023), meaning that violence associated with rice-producing areas is especially relevant to income generation for both farmers and the state.

Finally, while there are low levels of interstate conflict in Southeast Asia compared to South Asia or the Middle East, civil conflict and social unrest have been defining features of the region's politics (Caruso et al., 2015; Crost and Felter, 2020; Crost et al., 2020; Gatti et al., 2021). For example, the Philippines and Myanmar have experienced multiple insurgencies—ideology- or ethnicity-based—for nearly their entire period of independence. The protests and conflicts sparked by the Myanmar military's coup in 2021 have evolved into a full-scale civil war, while the Philippines has periodically experienced high levels of civil conflict. Thailand struggled with communist insurgency from the 1960s until the early 1980s, as well as spillover from Myanmar's insurgencies, while the Thai government continues to fight a low-level ethnic insurgency in southern Thailand. Since Indonesian President Suharto's fall in 1998, Indonesia has dealt with widespread protests, riots, civil conflict, and Islamist terrorism as well as ethnic insurgencies in Aceh and West Papua.

**A Theory of Harvest-Time Conflict**

Accepting there are many ways to look at the link between agriculture and conflict, we focus our theory on the relationship ignited from harvest-time windfalls, and we take as our starting point Collier and Hoeffler's (2004) theory of conflict as stemming from greed or grievance—the approach has been used amply to look at agriculture-related conflict (McGuirk and Burke, 2020; Hendrix and Haggard, 215; de Winne and Peersman, 2021).

The first driver of conflict, greed, suggests that perpetrators are more likely to engage in conflict when there is more at stake (McGuirk and Burke, 2020). The second driver, grievance, suggests that people protest the deterioration of their well-being relative to others or their own past (e.g., Hendrix and Haggard, 2015; de Winne and Peersman, 2021).

Our theory, which links harvest times and changes in levels of violence, accepts that both greed and grievance can be part of the dynamics of seasonal violence. We disaggregate motives for violence during harvest time (and accompanying changes in availability and affordability of food) by the type of actors and their relationship to the food supply and income from food production. In other words, the response of conflict actors to the harvest (and the concomitant agricultural surplus) is conditioned by the type of actor and the conflict in which they engage.

Sorting channels that motivate conflict is a challenging task, as the observed data are the fruits of those motivations rather than the motivations themselves. Differentiating the channels thus requires thinking through who might be engaging in what form of conflict as a result of the motivation derived from the channel, who or what would be the target of conflict or violence, and how the conflict would manifest itself in response to, in the present case, agriculture-related employment and income. Table 1 summarizes our theoretical expectations, on which we further elaborate below.

**Table 1. Actors and harvest-time conflict: Theoretical expectations**

| Forms of conflict | Type of actor | Likely motive | Harvest-time effect |
|---|---|---|---|
| Political violence (battles, violence against civilians) | Organized armed groups (state forces, rebel groups, militias) | Rapacity (+), Resentment (+) | Positive |
| Social unrest (protests, riots) | Unorganized actors (civilians) | Opportunity cost (−), Resentment (+) | Ambiguous |

Harvest may fuel conflict by presenting opportunities for perpetrators to extort agricultural surplus or incur damage to their opponents—the *rapacity channel*, as well as by temporarily altering the relative income distribution among groups of people involved in agricultural and non-agricultural activities—the *resentment channel*. Usually, resentment leading to social unrest occurs in the context of a *decrease* in crop production and food supply, either directly or intermediate through other factors. Climate extremes (resulting in lower food supply) can lead to deterioration in individuals' livelihoods, and a gap between actual welfare and perceived status (Fjelde, 2015), while a decrease in welfare for farmers can make already existent inequalities worse (Jones et al., 2017; Heslin, 2020). However, the logic of resentment can apply to increases in inequality as a result of seasonal agricultural surplus as well, as the haves are separated from the (non-agricultural producer) have nots (Ubilava et al., 2023). On the other hand, harvest may reduce conflict as potential conflict actors are busy harvesting, or the benefits from harvesting outweigh the costs incurred by forgoing this activity in favor of involvement in conflict—the *opportunity cost channel*.

A way to disentangle these channels is by focusing on the forms of conflict instigated by different types of actors. Organized actors—state forces, rebel groups, and militias—engage in battles with each other or in violence against civilians. Unorganized actors, including civilians, involve in protests staged against the government; they may also engage in riots. The logic of conflict, the relevance of agricultural income, and the pattern of seasonality are likely to be different and, in some ways, diametrically opposed, depending on the type of actor (and thus the form of conflict). The literature has found that different types of violence may have diverging outcomes in situations where conflict causes food insecurity, or vice versa. George and Adelaja (2022), for example, show that influxes of internally displaced persons due to armed conflict tend to decrease food consumption in host communities, while influxes due to communal violence tend to increase food consumption.

Specifically, because organized armed groups are carrying out planned, armed violence against adversaries, they are likely to engage in the violence both because they have grievances, and to extract resources for their fight. While they may have multiple motives (or channels through which motive can lead to a change in violence), in all cases the motives lead to an increase in violence during harvest times. On the other hand, because unorganized actors—notably civilians—are not otherwise engaged in active conflict, they do not need to forcibly extract surplus from the harvest, either to enrich themselves or to disadvantages their enemies. Indeed, civilians who are agricultural producers themselves, may have incentives to decrease unrest—protests and riots—during harvest.

We can summarize the foregoing discussion in two key arguments.

Argument 1. *Organized actors—armed groups engaged in a larger conflict—are likely to increase political violence during harvest time.*

In armed conflicts—both in interstate wars and civil wars—the combatant factions have incentives to create food insecurity while gaining material advantage over adversaries. This may include destruction of cropland, displacement of populations, and control of food, and denial of food or income to the adversaries or supporters of adversaries (Kemmerling et al., 2022). Indeed, timing attacks with harvests has been a time-tested tactic in warfare (Davis, 2002; Kreike, 2022).

Violence against civilians can be linked to the harvest-time income, and this relationship is expected to be positive. During the harvest season, an all but guaranteed increase in agricultural output makes farmers a lucrative target for perpetrators, which can amplify violence in croplands relative to non-agricultural areas (Mitra and Ray, 2014; Koren, 2018; McGuirk and Burke, 2020). This relationship, which aligns with the theory of greed manifested through the rapacity channel (Ubilava et al., 2023), can be further reinforced by the resentment channel, for example, if harvest-time change in relative income between farmers and non-farmers leads to communal violence.

By the same logic, organized armed actors (whether state forces, rebel groups, or militias) may increase their activities during the harvest season to maximize the damage they do in the larger conflict for several reasons. First, they may want to expropriate farmers' income, which is realized during the harvest season. Second, for farmers who do not support the insurgency or state, or who are on the sidelines, organized armed actors may want to harm their earning potential to minimize threats to themselves or to intimidate the farmers into joining them (Raleigh, 2012; Raleigh and Choi 2017). Third, rebels may strategically time their attacks to maximize the negative effect on the state, considering that the state is likely to derive revenue during the harvest season. This revenue is directly related to the state's ability to attack rebels (Fearon and Laitin, 2003). Linke and Ruether (2021) find evidence for this when they differentiate between 'violence to capture agriculture' and 'violence to destroy agriculture,' and posit that both motivations for violence would lead to an increase in violence during growing seasons. They find that this is indeed the case for violence initiated by both the government and rebel groups (in Syria, specifically).

In Southeast Asia, many attacks on civilians revolve around the theft of rice or the destruction of rice fields or rice storage facilities. In December 2013, for instance, the Myanmar military attacked civilian rice paddies in Kachin State during the harvest season.[1] In November 2021, in a clash between the Myanmar military and various People's Defense Forces (the armed groups linked with the anti-junta civilian government in internal exile), the military burned harvested rice fields in a village they suspected was loyal to the anti-state groups.[2]

---

[1] ACLED #9785083: "On 9 December 2013, in Mung Ding Pa, Kachin state, the Myanmar army shelled civilian rice fields."
[2] ACLED #8807873: "On 4 November 2021, west of Pekon township (Taunggyi district, Shan-South), military troops clashed with the joint forces of Pekon PDF, Moebye PDF, Loikaw PDF, Demoso PDF, the KNDF and the Karenni Army. Military troops fired artillery and torched, looted a nearby village and burned harvested rice in paddy fields according to Pekon PDF. At least 20 military troops were killed and a resistance fighter was injured."

Argument 2. *Unorganized actors—civilians not otherwise engaged in conflict—may increase or decrease social unrest during harvest time.*

As with political violence by organized armed groups, unorganized actors may also see resentment related to the harvest. Generally, the literature has seen the inverse relationship between food availability and affordability and protests: Higher food prices, the usual indicator for food scarcity, lead to decreases in well-being in populations, either relative to the past or to other groups, which results in resentment, often against the government if it is not perceived as addressing the problem. This resentment in turn leads to protests and riots (Kemmerling et al., 2022).

In rural areas, the mechanism that links food availability with conflict is nuanced. During harvest (that is, during times of plenty), while the grievances of farmers may decrease because of the realization of income, an increase in their income relative to others (rural non-farmers or urban dwellers) may lead to resentment between different groups (e.g., Panza and Swee, 2023), and an increase in social unrest between groups or against the government.

However, agricultural windfalls during harvest season can also increase the opportunity cost of protests and riots, which can occur through several pathways. First, those who are directly involved in agriculture may decrease their protest activities during harvest because they are busy harvesting. This relates to a second pathway in which the opportunity cost of protesting increases because of agricultural income realized at harvest time. This infusion of income makes protesting relatively less attractive, perhaps because there are fewer grievances against the government when would-be protesters are realizing income. While decreases in food access can enable mobilization around several other, often unrelated grievances (Heslin 2020), an *increase* in food access (during the harvest) may remove a means of mobilizing around pre-existing grievances, thus decreasing protests. A third possible pathway is the income *from* protesting becoming less competitive. In Indonesia, for instance, paid protesting is a long-standing means for political parties and civil society groups to pressure the government or to "send a message." Protesters, who typically are provided with a packed lunch ("nasi bungkus" in Indonesian, hence the term "nasi bungkus brigade") and a cash payment, often have only a tenuous interest in the cause of the protest (Andrews, 2017). Thus, protest campaigns may find fewer supporters during the harvest season.

The net differential effect of harvest season on social unrest by unorganized groups, as seen by the negative effect of opportunity costs and the positive effect of resentment, can be ambiguous. In Thailand, for instance, farmers staged a spate of protests in 2014 in relation to a rice purchasing scheme in which the Thai government was supposed to have paid subsidies for rice production, but the payments were either delayed or absent (Mohanty, 2012). These protests did not happen during the rice harvest season (which is December in Thailand), but several months later, presumably when farmers' grievances increased, and the opportunity cost of protesting decreased. Of interest here is that Thai farmers' grievances were against both government (for not paying subsidies) and anti-government groups (for supposedly blocking the government from paying subsidies).[3]

**Data**

Our data come from multiple sources. For social conflict, we use the Armed Conflict Location & Event Data (ACLED) compiled by Raleigh et al. (2010). For rice cropland cover and irrigation, we use data from IFPRI (2019), and for harvest calendars, we use data from Sacks et al. (2010). For precipitation, we use the CPC Global Unified Gauge-Based Analysis of Daily Precipitation data provided by the National Oceanic and Atmospheric Administration's Physical Sciences Laboratory.[4] For city population, we use World Cities Database.[5] In what follows, we describe these data in detail.

---

[3] ACLED #7908847: "Farmers in the province of Phichit staged a protest against the anti-government movement, blasting its attempt to block the government's efforts to secure funds for the rice pledging program. Hundreds of farmers gathered at a major intersection to express their opposition to the People's Democracy Reform Committee (PDRC), who they believed have been blocking the government's attempt to pay rice farmers for rice pledged under the pledging program."

*Social conflict*

The ACLED Project (Raleigh et al., 2010) presents spatially and temporally granular data on conflict incidents that are split into six categories: battles between organized armed groups (which includes state forces, rebel groups, and militias), explosions/remote violence, strategic developments, violence against civilians by organized armed groups, and protests and riots representing forms of public disorder. In our analysis, we combine battles and explosions/remote violence into a single form of conflict and exclude strategic developments, as they are not comparable across countries or over time (Raleigh et al., 2010).

The distinction between organized actors and the violence they commit and unorganized actors who engage in protests and riots comes from how the ACLED categorizes actors and incidents. Both battles and violence against civilians, in ACLED's formulation, require perpetrators to be organized into armed groups. Battles are "a violent interaction between two organized armed groups at a particular time and location" (ACLED 2023, p. 11), while violence against civilians comprise "violent events where an organized armed group inflicts violence upon unarmed non-combatants…The perpetrators of such acts include state forces and their affiliates, rebels, militias, and external/other forces." (ACLED 2023, p. 17)

In contrast, protests and riots are committed by civilians or, at the very least, unorganized actors. By definition, "[p]rotesters are peaceful demonstrators" (ACLED 2023, p. 29), and protesters are inherently civilians, inasmuch as "civilians are unarmed and cannot engage in political violence" (ACLED 2023, p. 17). For their part, rioters are unorganized. They are "loosely assembled groups of individuals or mobs without inherent organization, engage in violence while participating in demonstrations or engage in violence that is spontaneous… While the activity of rioters, by definition, falls outside the remit of an organized armed group, rioters may sometimes be armed and/or organized in a spontaneous or atomic manner." (ACLED 2023, p. 28). While riots may come from protests instigated by organized groups, violence is categorized as a riot if it is committed by loosely assembled, unorganized individuals.

---

[4] https://psl.noaa.gov
[5] https://simplemaps.com/data/world-cities

The dataset covers all major countries in Southeast Asia with an agricultural sector over a relatively short period of time, from 2010 onward for most countries (Cambodia, Laos, Myanmar, Thailand, and Vietnam), except for Indonesia (from 2015 onward), the Philippines (from 2016 onward), and Malaysia (from 2018 onward). We exclude Brunei, Singapore, and Timor-Leste because they are small and/or not agricultural countries, and because of the very short coverage for these countries (from 2020 onward).

Our study covers more than 86,000 unique incidents. This excludes incidents for which exact locations are unknown, and they are thus arbitrarily attributed to the nearest known site, typically a provincial capital (such locations are recorded with geo-precision code 3 in the database). Figure 1 presents a map of incidents across three distinct conflict categories aggregated at the level of one-degree cells. The map also features a selected set of large cities in the region.

From this map, it becomes apparent that conflict occurs across much of Southeast Asia. Within the region, some countries are more prone to conflict than others. There is notable spatial dependence in the prevalence of different forms of conflict, and while generally observed in cities, where most people reside, conflict is not necessarily or exclusively a city phenomenon.

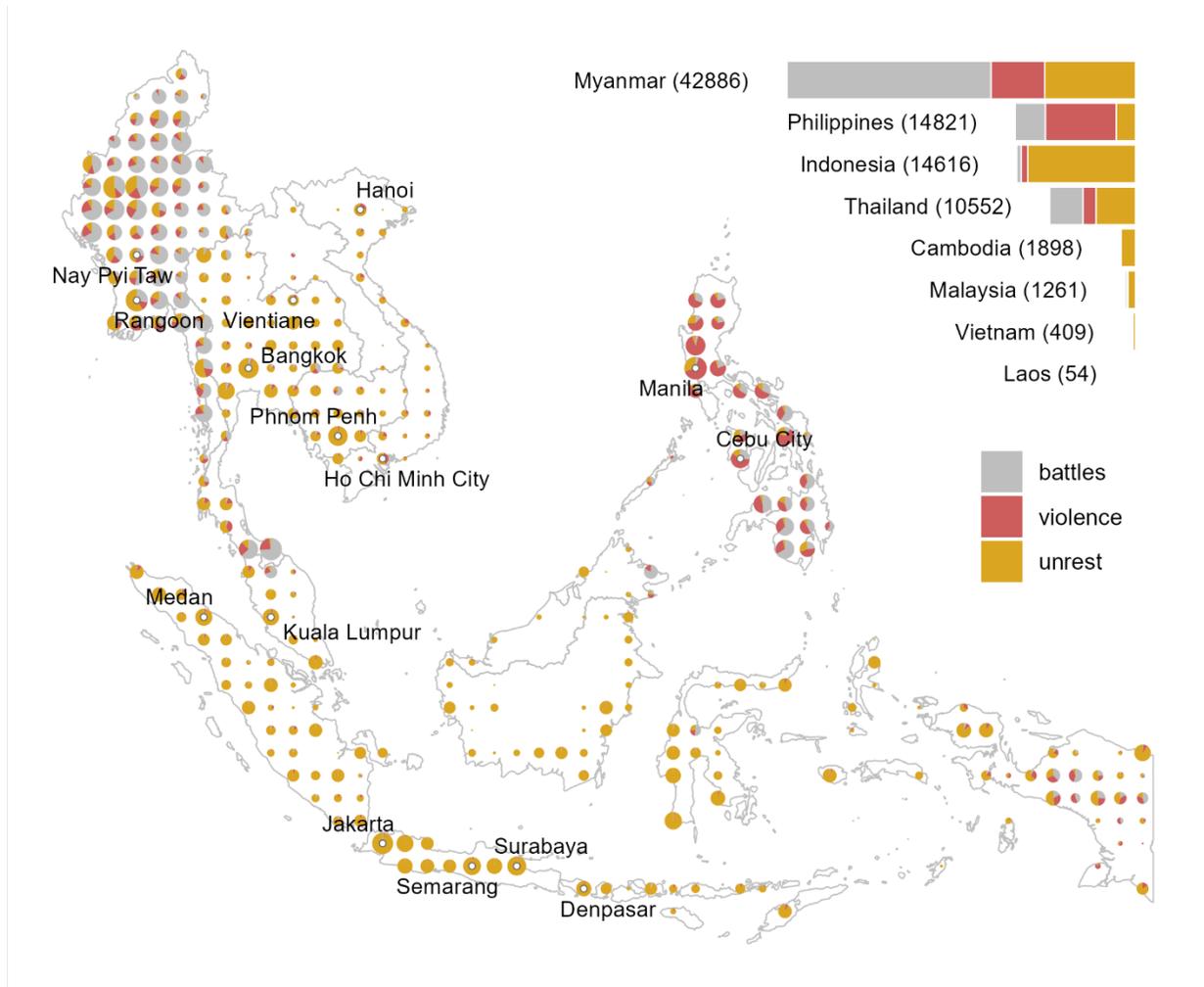

**Figure 1: Geographic distribution of social conflict (2010–2023)**

Note: The data are for Cambodia, Indonesia (2015–2023), Laos, Malaysia (2018–2023), Myanmar, the Philippines (2016–2023), Thailand, and Vietnam. The size of the dots is proportional to the combined number of incidents in a cell. Protests and riots are combined into a single unrest category for illustration purposes. The empty circles that overlay some of the dots identify "urban" cells, defined as those that include a capital city, a city with a population of at least one million, or a total within-cell population of at least two-and-half million. The map also includes labels for capital cities and large metropolitan areas (with city population of at least two-and-half million). The data sources are ACLED and the World Cities Database.

The proportions of forms of conflict also vary by country: Myanmar saw high levels of unrest and a high number of battles, while the Philippines saw little unrest but high levels of violence against civilians. By comparison, Indonesia saw almost no battles or violence against civilians, but high levels of unrest, and Thailand was split relatively equally between unrest and battles. Figure 2 shows the time series of the four forms of conflict considered.

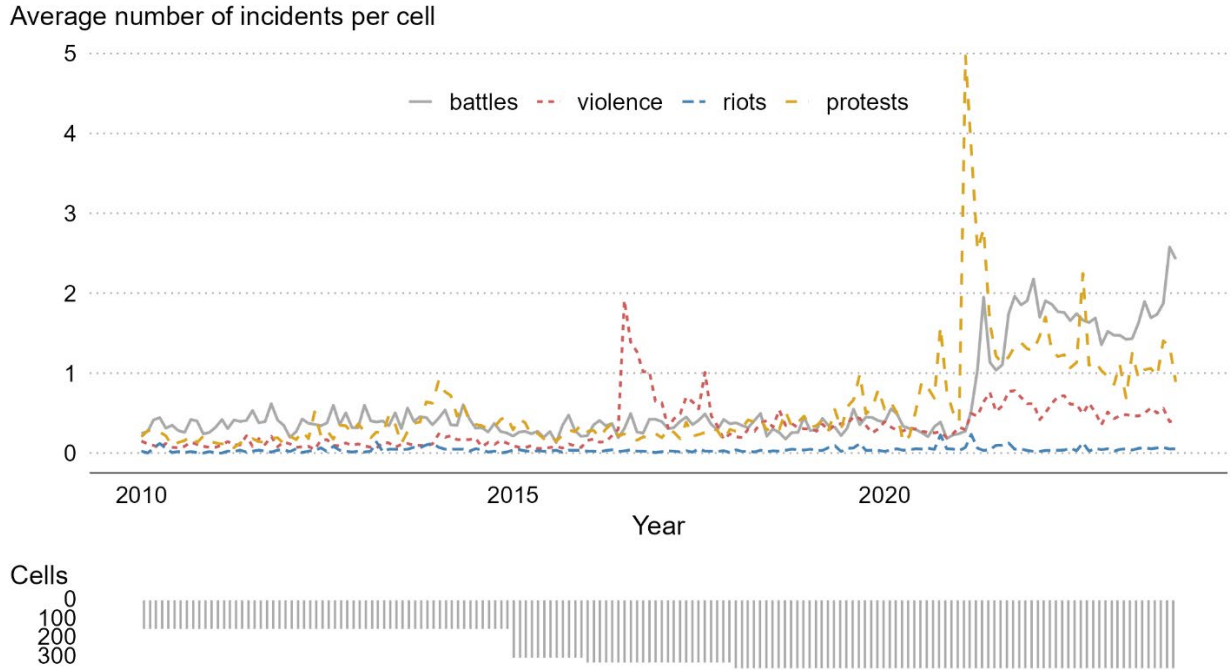

**Figure 2: Dynamics of social conflict by event type**

Note: The time series are monthly rates of conflict incidents per cell across available cells for a given period. The number of cells (bottom panel) increased progressively as Indonesia (2015), the Philippines (2016), and Malaysia (2018) were added to the dataset. The data source is ACLED.

The figure shows no apparent trend across the forms of conflict, but there is a notable increase in almost all forms of conflict, except riots, from 2021 onward—a change largely driven by the Myanmar Civil War. Despite general co-movement, there are periods when a rise in one form of conflict is not accompanied by other forms of conflict, suggesting that the root causes and mechanisms of different forms of conflict vary.

*Rice harvest season*

We focus on rice, which is by far the most dominant cereal—both in terms of production and consumption—across Southeast Asia. The harvest can span several months. We define the period from the month when the harvest starts to the month when the harvest ends as the harvest season. We define the midpoint of the harvest season as the harvest month. In instances where a crop is grown over two seasons, we use the main season to identify the crop year. We keep the area of the cropland and the harvest season fixed over the study period. We do so due to data limitations, but also to mitigate the issue of reverse causality from conflict to harvest.

Figure 3 presents the geographic prevalence of rice production and the harvest months at the level of one-degree cells. The map makes it apparent that there is some variation in the timing of the main harvest season, with March being the dominant month. There is also considerable within-country variation in cropland area, but hardly any within-country variation in the harvest month.

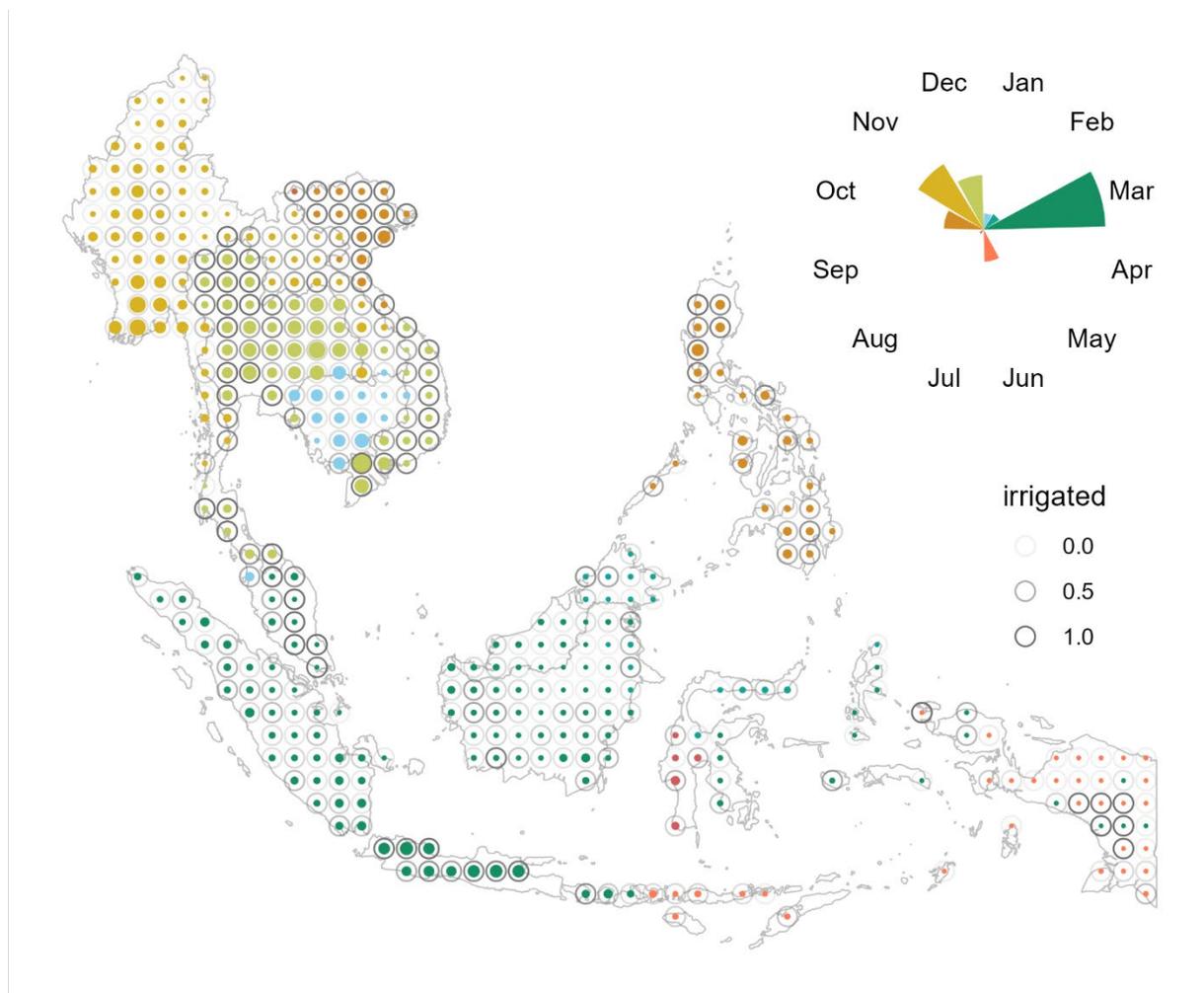

**Figure 3: Geographic distribution of rice harvest months**

Note: The size of the dots is proportional to the area devoted to rice production, and the radial bars indicate the count of cells that fall within a given harvest month. The empty circles that overlay the dots identify irrigated cells, opacity indicating the higher proportion of irrigated land. The data sources are IFPRI (2019) and Sacks et al. (2010).

*Auxiliary data*

We use rainfall, one of the key inputs in rice production, it to test the mechanism related to the year-to-year change in rice production and thus its abundance at harvest time, in the crop-producing cells. In the analysis we apply monthly average growing season precipitation, which we obtain by dividing the aggregate observed precipitation in one-degree grid cells—the spatial unit of measurement in this study—by the number of months in the crop growing season, i.e., the months from the start of the planting to the start of the harvest season. Appendix Figure B1 presents a histogram of the monthly average growing season precipitation across the cells and years used in the analysis.

We use information on irrigation to differentiate regions with relatively more developed infrastructure from regions with minimal infrastructure for rice production. Empty circles on Figure 3 show the cells with irrigated land, high-opacity circles indicating the cells with high proportion of irrigated land. Appendix Figure B2 presents a histogram of the proportion of irrigated rice across all considered locations. Cells with larger cropland areas tend to be irrigated, although the irrigation prevalence also appears to be a country-specific phenomenon.

The state—both the nature of the regime, and state capacity—may serve as an intermediating factor between agricultural windfalls and conflict, although most research looks at food *in*security and conflict, rather than windfalls. Democracies tend to see more social unrest during periods of high food prices than autocracies (Hendrix and Haggard 2015). Food insecurity is more likely to lead to conflict in politically vulnerable states (Jones et al, 2017). To get some measure of state presence in an area, we use the information on population size to categorize cells into urban and rural, with the assumption that urban areas may have more state presence than rural areas. We consider a cell to be urban if it contains the capital city or a large city (with a population size of at least one million), or if the population size within the cell, which may include multiple cities, is at least two-and-half millions. Empty circles on Figure 1 indicate these cells.

*Descriptive statistics*

In Table 2, we summarize some of the data's key features. Battles combined with explosions/remote violence, labeled as "Battles," violence against civilians, labelled as "Violence" and protests represent the three most prevalent forms of conflict. The least prevalent event type is riots—a violent version of protests that shares elements of other armed forms of social conflict. The average size of the land used for rice production is approximately 78,000 hectares, which is approximately 6.5% of the cell (measured near the equator).

**Table 2: Descriptive statistics**

|  | Sum | Mean | SD | Min | Max | Incidence |
|---|---|---|---|---|---|---|
|  | *Unit of observation: Cell-year-months (49,236)* | | | | | |
| All events | 86,083 | 1.748 | 7.759 | 0 | 349 | 0.266 |
| Battles | 33,979 | 0.690 | 4.223 | 0 | 136 | 0.100 |
| Violence | 17,998 | 0.366 | 2.681 | 0 | 164 | 0.094 |
| Riots | 2,603 | 0.053 | 0.439 | 0 | 42 | 0.034 |
| Protests | 31,503 | 0.640 | 3.731 | 0 | 267 | 0.163 |
|  | *Unit of observation: Cells (376)* | | | | | |
| Rice cropland area (100,000 ha) |  | 0.781 | 1.255 | 0 | 9.087 |  |
| Irrigated |  | 0.348 | 0.738 | 0 | 7.786 |  |
| Rainfed |  | 0.433 | 0.858 | 0 | 5.759 |  |

Note: The conflict data are from the ACLED Project (Raleigh et al., 2010) and cover eight countries over the 14-year period from 2010 to 2023, except for Indonesia (2015–2023), Malaysia (2018–2023), and the Philippines (2016–2023). The other countries are Cambodia, Laos, Myanmar, Thailand, and Vietnam. *All events* contain the four forms of conflict, wherein *Battles* combine battles and explosions/remote violence (as defined by the ACLED Project). *Incidence* denotes the proportion of units with at least one conflict incident. The data on rice croplands, which are fixed at levels circa 2010, are from IFPRI (2019).

**Estimation, Identification, and Interpretation**

We denote location, a one-degree grid cell, with subscript $i$, country with subscript $c$, year with subscript $t$, and month with subscript $m$. Henceforth, we refer to the year-month as the period of observation. Thus, the units of analysis are observations across 376 unique grid cells and, in most cases, over 168 periods from January 2010 to December 2023. The level of spatial aggregation—one-degree cells that measure approximately 110×110 km near the equator—is coarse enough to ensure that there are enough units with at least one conflict incident. This level of aggregation is sufficiently granular to not sabotage within-country variation in conflict incidents.

Our main econometric specification is given in a fixed effects setting as follows:

$$Conflict_{itm} = \beta Cropland_i \times Harvest_{im} + \mu_i + \lambda_{ct} + \gamma_{tm} + \theta' Z_{itm} + \varepsilon_{itm}, \qquad (1)$$

where the outcome variable, $Conflict_{itm}$, is the number of incidents in cell *i* in month *m* of year *t*. The treatment variable is the product of $Cropland_i$, which is the cropland area used for rice production (IFPRI, 2019) measured in 10,000 hectares, and $Harvest_{im}$, which is the binary variable that equals one when the period of observation falls on the harvest season in the cell and equals zero otherwise (Sacks et al., 2010). $\mu_i$ is a cell fixed effect, $\lambda_{ct}$ is a country-year fixed effect, and $\gamma_{tm}$ is a year-month fixed effect. $Z_{itm}$ is a set of controls—in most specifications the contemporaneous rainfall—that vary across locations and over time. $\varepsilon_{itm}$ is the error term.

The identifying assumption in Equation (1) is that the treatment variable is exogenous to conflict. This assumption may seem tenuous because conflict may affect production through factors such as abandoned plots, missed or mistimed harvests, and planting seasons. As a result, lower agricultural output may be a consequence of the change in conflict rather than its cause. However, in our study, we do not apply production data that would vary yearly and instead use cropland area and harvest months, which are location-specific and fixed over time. Such an approach, admittedly driven by data limitations, mitigates the issue of reverse causality.

We control for a set of variables in the regression. We include cell fixed effects to capture any time-invariant determinants of conflict (e.g., distance to roads, cities, or state borders), country-year fixed effects to control for any country-specific trends in the data (e.g., large-scale political turmoil in election years), and year-month fixed effects to capture common time-varying events observed in the region (e.g., global financial crises, large-scale climatic shocks, and possible trends or seasonality in the quality of data collection/reporting). We also include contemporaneous rainfall, which varies over time and across space, in the regression to address, at least to an extent, the remaining endogeneity issues. This allows us to account for the direct impact of weather on conflict; for example, if excessive rainfall reduces the mobility of troops or makes protests and demonstrations somewhat untenable. Finally, in the robustness checks, we vary the fixed effects to obtain a better sense of potential threats to our identification strategy.

The estimated coefficient, $\hat{\beta}$, reflects the harvest-time change in conflict in the cells with sizeable croplands relative to the cells with small or no croplands. For example, a positive value of this coefficient would imply that in croplands, there is more conflict during the months of harvest compared to the rest of the year. To help interpret the effect, we evaluate it at the average size of the cropland and the expected number of conflict incidents.

**Results and Discussion**

Table 3 presents the baseline results of this study using all available data. During the harvest season, we estimate an increase in forms of political violence and a decrease in forms of social unrest. All coefficients, except that for riots, are statistically significantly different from zero at the 5% level. Focusing on the statistically significant effects, relative to the baseline conflict and evaluated at an average cropland, which is measured at approximately 9,000 ha within a cell, the estimated coefficients translate to a 6.7% increase in battles, a 9.1% increase in violence against civilians, and a 9.2% decrease in protests during the harvest season.

The harvest-time increase in political violence supports rapacity as the dominant channel driving agrarian conflict in times of a relative food abundance. The harvest-time decrease in social unrest is consistent with the opportunity cost channel.

**Table 3: Harvest-time conflict in the croplands of Southeast Asia**

|  | All events | Battles | Violence | Riots | Protests |
|---|---|---|---|---|---|
| *Unbalanced panel: All countries, all years* | | | | | |
| Cropland×Harvest | 0.018 | 0.050** | 0.036*** | -0.005 | -0.064** |
|  | (0.036) | (0.021) | (0.010) | (0.003) | (0.030) |
| Obs. | 49,236 | 49,236 | 49,236 | 49,236 | 49,236 |
| $R^2$ | 0.403 | 0.381 | 0.371 | 0.160 | 0.319 |
| *The magnitude of the estimated effect relative to the baseline conflict in the average cropland (%)* | | | | | |
| *Baseline conflict* | 1.75 | 0.69 | 0.37 | 0.05 | 0.64 |
| *Average cropland* | 0.91 | 0.91 | 0.91 | 0.91 | 0.91 |
| Cropland×Harvest | 0.9 | 6.7** | 9.1*** | -7.8 | -9.2** |
|  | (1.9) | (2.7) | (2.5) | (5.7) | (4.2) |

Note: The outcome variable is the number of conflict incidents in a cell in a year-month, and the treatment variable is the cropland area (10,000 ha) interacted with the indicator for the harvest season. The "All events" column combines all forms of conflict, and the "Battles" column battles and explosions/remote violence. The remaining columns represent the separate event types as labeled. All regressions include cell, country-year, and year-month fixed effects and control for contemporaneous rainfall. The values in parentheses are standard errors adjusted to clustering at the cell level; ***, **, and * denote 0.01, 0.05, and 0.10 statistical significance levels. The magnitudes of the effect, presented in percentage terms, are calculated as $\hat{\beta} \times \bar{a}/\bar{c} \times 100\%$, where $\hat{\beta}$ is the parameter estimate, $\bar{a}$ is the average cropland size, and $\bar{c}$ is the baseline conflict.

*Robustness to data subsetting and alternative specifications*

Before investigating possible mechanisms, we check that the results are not sensitive to data subsetting or varying model specifications.

First, we re-estimate the baseline model using balanced panels that either cover all eight countries but only include the years 2018 to 2023 (short balanced panel) or cover all 14 years but exclude Indonesia, Malaysia, and the Philippines (long balanced panel). Appendix Tables A1 and A2 present the regression results. When we apply the short balanced panel, the estimated coefficients for each of the four forms of conflict are largely comparable to those of the main results. Indeed, all coefficients are statistically significant at the 5% level in this instance. When we apply the long balanced panel, most of the estimated coefficients, with the exception of that for violence against civilians, become statistically indistinguishable from zero.

Second, we re-estimate the baseline model by omitting one country at a time and one year at a time. Appendix Figures B3 and B4 present the estimated effects, which are largely robust to omitting subsets of data from the analysis, with two notable exceptions. The harvest-time effect on violence against civilians while similar in magnitude, becomes statistically indistinguishable from zero when we omit Myanmar from the data. This same effect attenuates to zero when we omit 2016 or 2021 from the data. In all instances, we are discarding a considerable proportion of incidents, specifically of this form of conflict. So, the observed sensitivity is not surprising. This discrepancy, moreover, may also be attributed, at least to an extent, to escalation of conflict in Myanmar, specifically the civil war that followed the coup d'état in early 2021, as well as to behavioral and policy responses to the COVID-19 pandemic (e.g., Bloem and Salemi, 2021).

Third, we re-estimate the parameters using a set of alternative model specifications to ensure that our main results are not driven by our choice of fixed effects or that the inference is not affected by our choice of error clustering. We summarize these results in the specification chart presented in the Appendix Figure B5. The point estimates are not sensitive to different combinations of fixed effects. The evidently most robust estimate of the statistically significant increase in violence against civilians during the harvest season stands in most instances, except when we cluster the standard errors at the country level.

Fourth, we re-estimate the parameters using half-degree grid cells, which measure approximately 55×55 km, to ensure that the results are not sensitive to our choice of the level of spatial aggregation (which is one-degree cells). Appendix Table A3 present this regression results, which accord with the main results of this study.

Finally, we perform a falsification test to ensure that the estimated effects are not merely coincidental. Specifically, we shuffle and randomly re-assign the observed harvest seasons to different locations and re-estimate the baseline regression, repeating this 100 times. On average, we would expect no significant effect. Appendix Figure B6 confirms this. Apart from a few statistically significant estimates of the impact—likely stemming from a mere happenstance—we observe no substantial impact when the "wrong" harvest seasons are randomly assigned to the croplands.

*Treatment heterogeneity*

To the extent that the assumed treatment homogeneity may be camouflaging otherwise heterogeneous effects, we examine harvest-time forms of conflict across subgroups of locations sorted by some cell-specific characteristics.

Rainfed vs irrigated cells. First, we interact the treatment variable with the proportion of irrigated land in the cell (see Figure 3 and Appendix Figure B2). Figure 4 presents the estimated effects relative to the baseline conflict in the average cropland across the range of rainfed and irrigated croplands (Appendix Table A4 presents the underlying regression results).

The main takeaway is that harvest-time changes in conflict occur primarily in rainfed croplands. In general, irrigated rice is of a higher yield and is often commercially produced, as opposed to rainfed rice, which is of a lower yield and is typically produced at subsistence levels. Therefore, different types of farmers are likely to be involved in these two production practices. That is, irrigation likely serves as a catch-all variable for more developed/commercial areas with relatively more stable institutions, which makes them less prone to conflict on average.

Rural vs urban cells. Next, we interact the treatment variable with rural and urban indicator variables. Because social conflicts are most prevalent in places with high population densities, we check whether there is a qualitatively meaningful discrepancy in the seasonal conflict between urban and rural locations. We define a cell as urban if it contains the capital city or a city with a population size of at least one million or if the cell's population size exceeds two-and-half millions. There are 47 such cells, and more incidents of conflict occur in these cells than in other cells. Recall that the majority of these urban cells tend to have sizable croplands (Figure 3). Figure 5 presents the estimated effects from these regressions (Appendix Table A5 shows the parameter estimates).

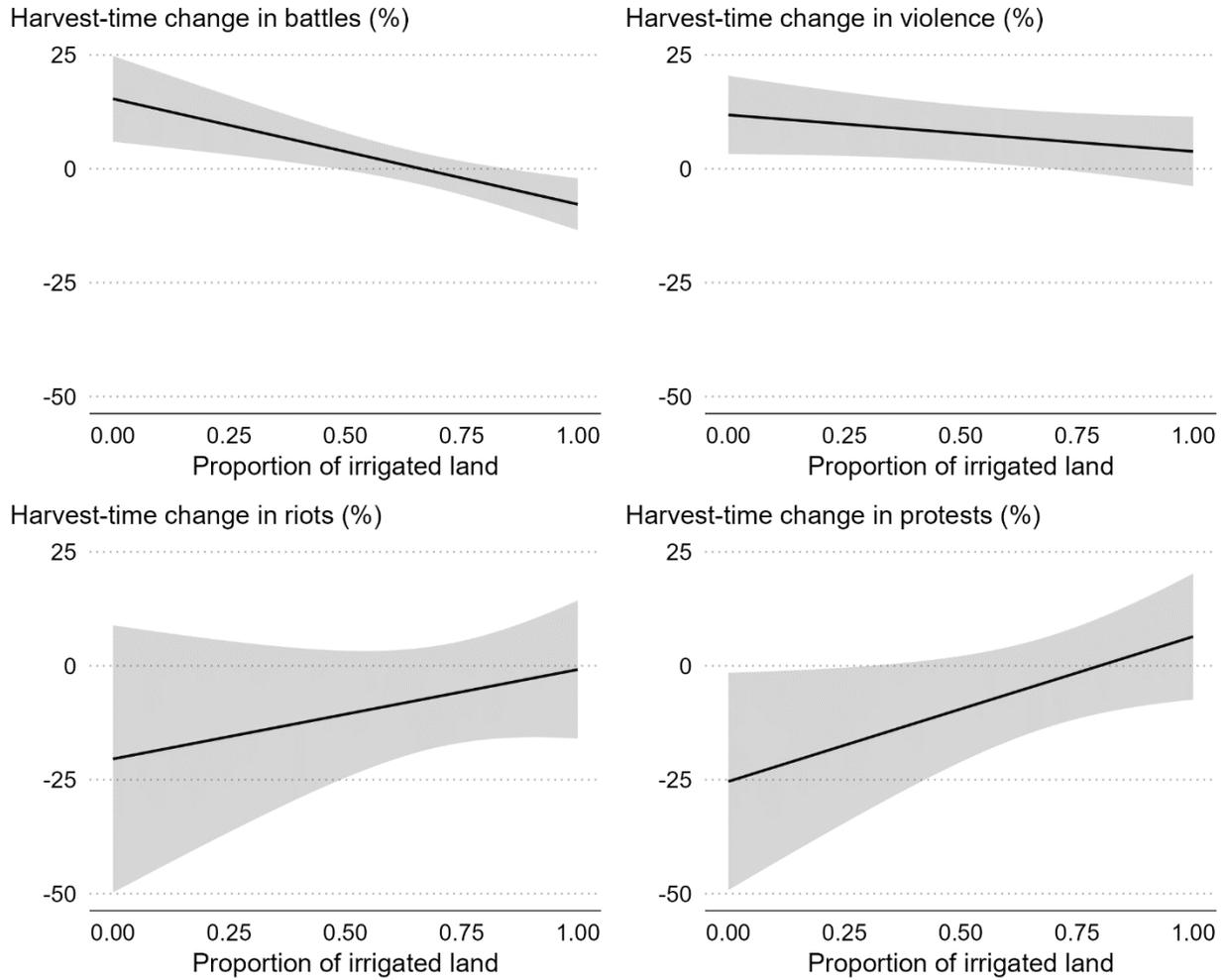

**Figure 4: Harvest-time conflict in rainfed versus irrigated cells**

Note: Solid lines show harvest-time percent change in conflict relative to the baseline conflict in the average cropland across the range of rainfed and irrigated croplands. The shaded bands show 95% confidence intervals based on standard errors adjusted to clustering at the cell level.

The results suggest that any harvest-time change in conflict, particularly an increase in violence against civilians and a decrease in protests, occurs only in rural areas. This finding supports the theory that civil conflict and armed violence are generally more common in peripheries, where either state policing is largely absent or where insurgents are typically present (e.g., Buhaug and Rød, 2006).

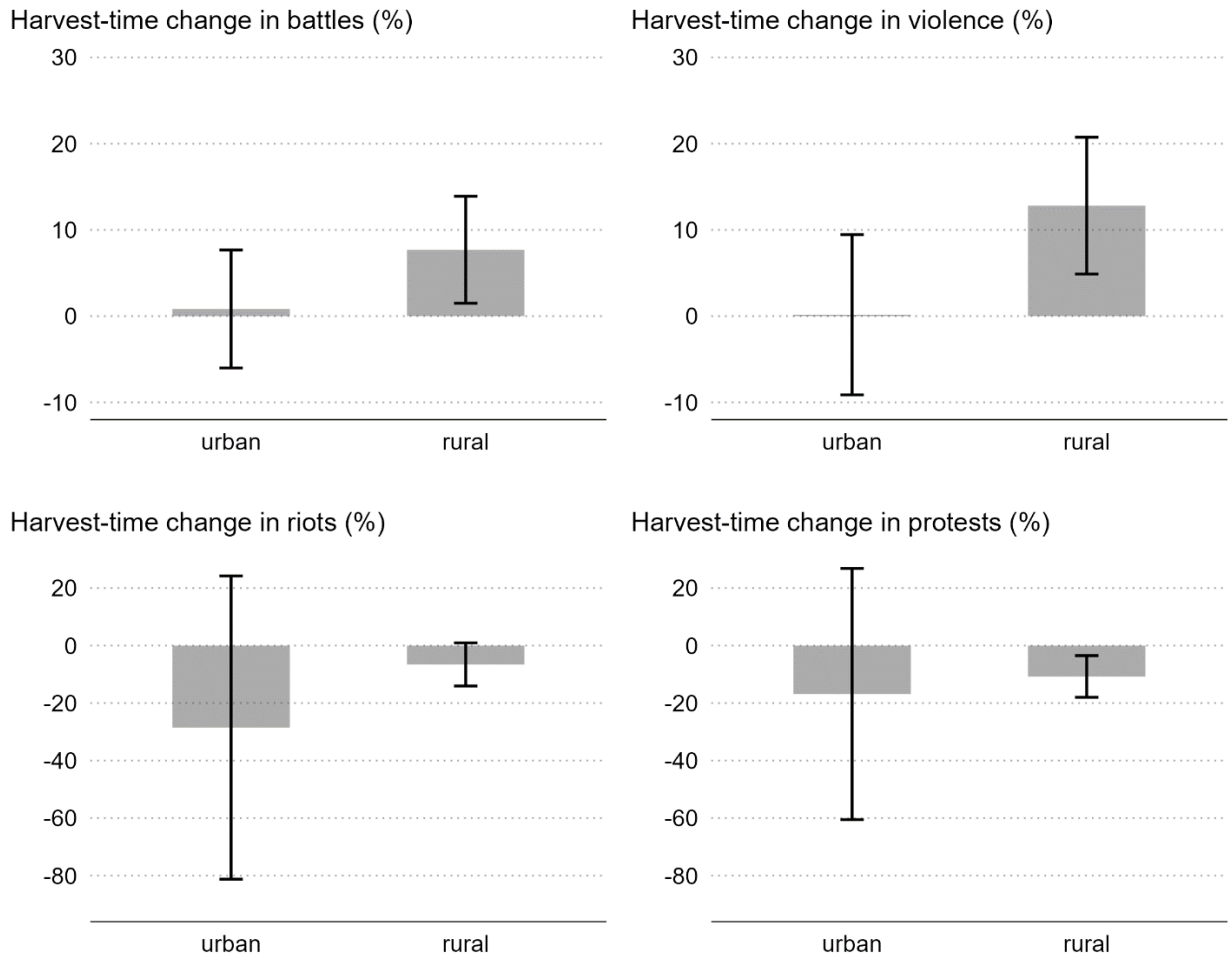

**Figure 5: Harvest-time conflict in rural versus urban cells**

Note: Gray bars show harvest-time percent change in conflict relative to the baseline conflict in the average cropland in urban and rural cells. The error-bars show 95% confidence intervals based on standard errors adjusted to clustering at the cell level.

*Testing the mechanisms*

The foregoing results bring context to motives that likely factor in the harvest-time conflict, namely rapacity in the case of political violence and opportunity cost in the case of social unrest. In what follows, we present additional regression results that will help us examine the mechanisms that manifest in these observed effects.

Growing season rainfall. First, we leverage the annual variation in the rainfall over the crop growing season to identify the effect of harvest on conflict. Both scarce or excessive rainfall can plausibly reduce crop yields due to drought or flooding (Rosenzweig et al. 2002; Hendrix and Salehyan, 2012; Wang et al., 2022; Chen and Wang, 2023; Fu et al., 2023). We compare the harvest-time changes in conflict across growing seasons ranging from scarce or excessive monthly levels of rainfall. We do so by interacting the treatment variable with linear and quadratic terms of the monthly average growing season rainfall (measured in 100mm). Figure 6 presents the differential effects of growing season rainfall on the forms of conflict at harvest (Appendix Table A6 shows the underlying regression results).

Several patterns emerge. First, we observe a harvest-time increase in violence against civilians after a growing season with average (presumably close to optimal) levels of monthly growing season rainfall, which accords with our proposed linking of the harvest-time effect to the rapacity and resentment channels. Second, we observe a harvest-time increase in battles after a growing season with scarce rainfall. This contradicts the idea that rapacity or resentment are the driving forces of this form of conflict but accords with the existing literature that links adverse weather conditions, drought in particular, with increase risks of conflict in rural or agrarian societies (e.g., von Uexkull et al., 2016). An alternative mechanism might be at play here. Because dry conditions enable troops to engage in a military combat, the mechanism that links growing season weather with harvest-time battles is likely a manifestation of a direct effect rather than that mediated via agriculture.

In the case of social unrest, and specifically in the case of protests, in good crop years, all three pathways (of employment, income, and alternative income) of the opportunity cost channel point in the same direction, but there is an offsetting effect from the resentment channel. In bad crop years, two of the three pathways of the opportunity cost channel likely reverse: while farmers will harvest the crop regardless of its quality, lower yields will reduce the opportunity costs of protesting associated with income from farming or income from protesting. But the offsetting effect from the resentment channel will also vanish due to a smaller harvest-time change in within-cell income inequality between farmers and non-farmers. It is generally difficult, and in the present setting impossible, to disentangle these effects. But we can argue which channels/pathways dominate the effect by observing the combined effects in good and bad crop years. The inverted U-shaped relationship between growing season rainfall and harvest-time social unrest, protests in particular, seems to suggest that to the employment-related opportunity cost channel and the resentment channel are the two likely drivers of the harvest-time changes in social unrest: in good years, the effect stemming from the resentment channel nearly fully offsets the opportunity cost effect. This offsetting does not happen in years with relatively poor harvest.

Growing season conflict. Next, we investigate whether the harvest-time change in conflict is a byproduct of a lingering military activity in the area. A conflict that coincides with the crop growing season, may have direct and indirect effects on harvest-time conflict. The direct effect, which entails some form of path dependence, is likely to be positive. The indirect effect, manifested through the agricultural channel, is more likely to be negative, as conflict will likely deter farmers from their usual activities, and thus, lower agricultural output (Olsen et al., 2021).[6] This, in turn, may mute the rapacity and resentment channels associated with the harvest-related political violence.

We obtain a cell-specific indicator variable of conflict that takes a value of one if at least one incident of any form of conflict occurred during the crop growing season in the cell. We then interact this indicator variable with the treatment variable to estimate harvest-time changes in the forms of conflict in the wake of the peaceful vs. conflict-prone crop growing season. Figure 7 presents the effects of growing season rainfall on the forms of conflict at harvest (Appendix Table

---

[6] Although, the effect of conflict and agricultural activities, which is complex and context-specific, may as well be positive (e.g., Baumann and Kuemmerle, 2016; Eklund et al., 2017)

A7 shows the underlying regression results). In addition to the general positive effect on all forms of postharvest conflict (see Appendix Table A7), conflict incidence during the crop-growing season has a positive differential effect on harvest-time change in political violence and a negative differential effect on harvest-time changes in social unrest. Among these effects, only that associated with violence against civilians is statistically significantly different from zero.

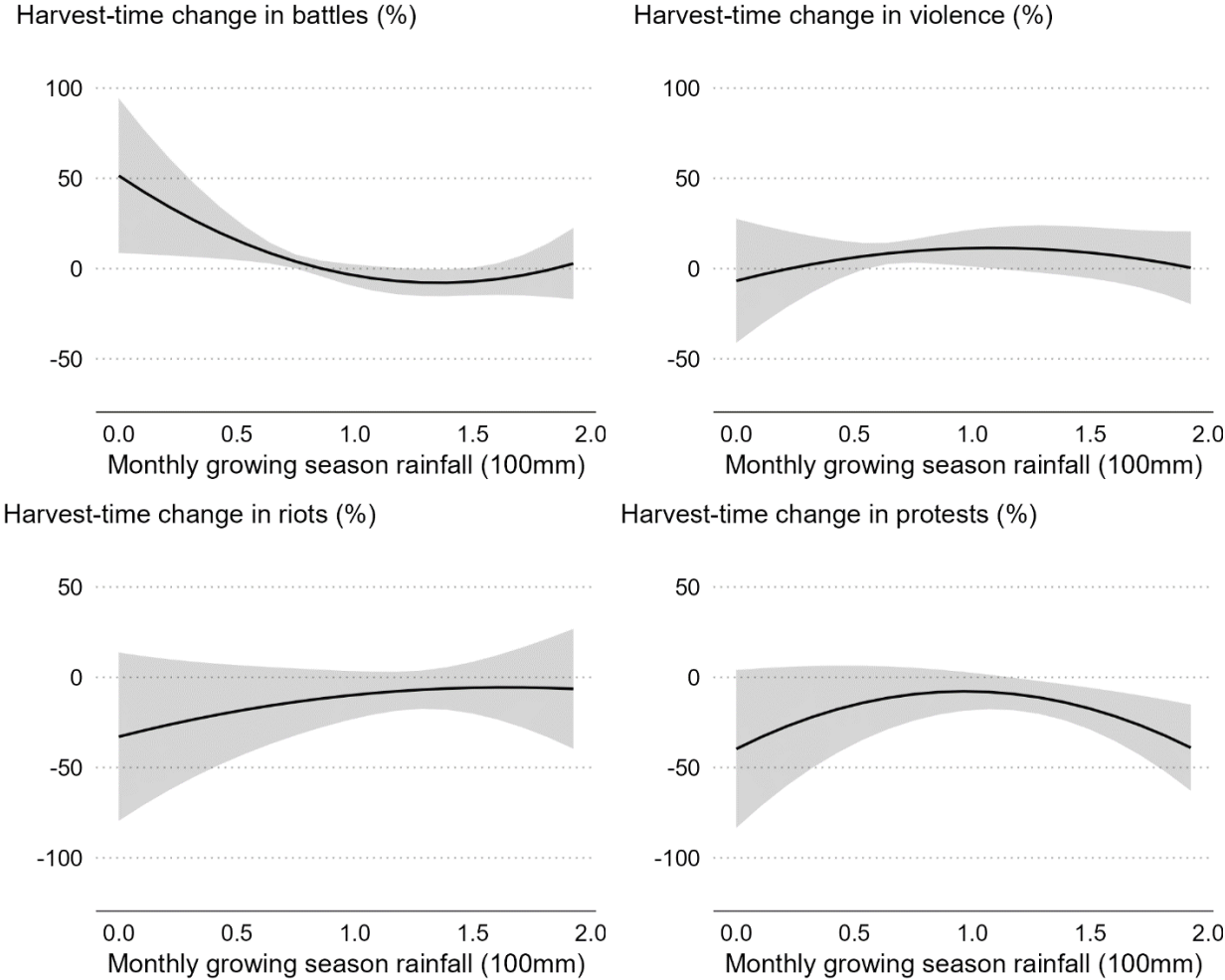

**Figure 6: Growing season rainfall and the harvest-time conflict**

Note: Solid lines show harvest-time percent change in conflict relative to the baseline conflict in the average cropland across the range of monthly average growing season rainfall. The shaded bands show 95% confidence intervals based on standard errors adjusted to clustering at the cell level.

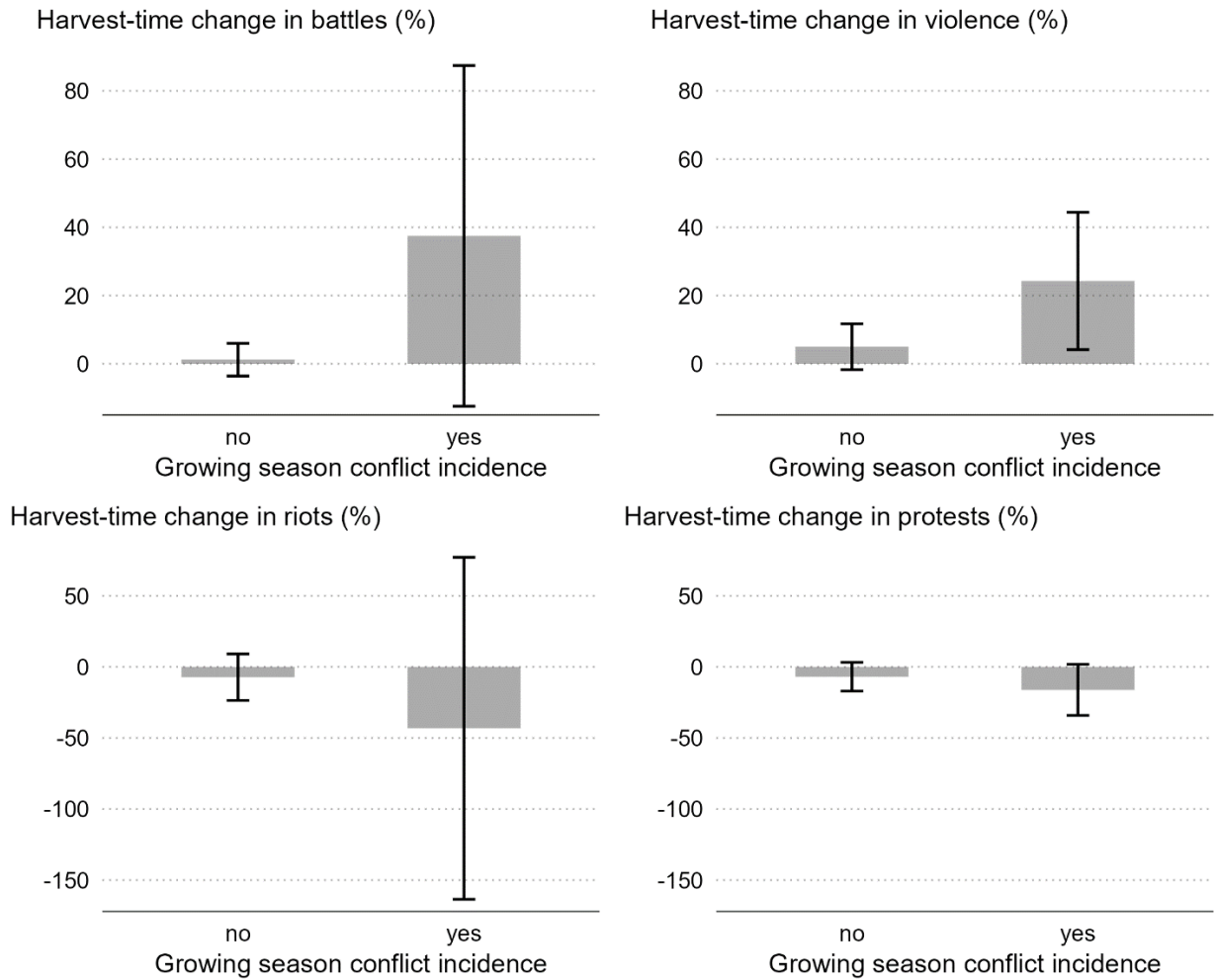

**Figure 7: Growing season conflict and the harvest-time conflict**

Note: Gray bars show harvest-time percent change in conflict relative to the baseline conflict in the average cropland in cells that were or were not exposed to the growing season conflict. The error-bars show 95% confidence intervals based on standard errors adjusted to clustering at the cell level.

Harvest-related seasonal political violence can be viewed as a byproduct of broader political conflict. In the case of political violence, this is consistent with the 'living off the land' theory, suggesting that in times of war, any co-optation between fighters and farmers breaks down, leading to more violence (Koren & Bagozzi, 2017). In the case of social unrest, the effect of growing season conflict mimics the effect of the growing season adverse weather: the harvest-time reduction in protests and riots is more pronounced when, for whatever reason, agricultural production is subpar.

**Conclusion**

How crucial is the harvest season in explaining the agricultural roots of social conflict? We address this question using 14 years of monthly data on social conflict from 376 cells with one-degree resolution across eight countries in Southeast Asia—a region where rice agriculture is one of the main sectors for employment and income. To better understand the pathways between harvest and conflict, we adopted a disaggregated approach. Different forms of conflict are typically carried out by different groups of people for different reasons: *political violence* (armed clashes between conflict actors and violence against civilians) usually involves the state, allied militias, or anti-state rebel groups, while *social unrest* (protests and riots) tends to be carried out by unorganized actors, often against the state or other groups.

So, instead of resolving the debate over the motives and mechanisms of resource-related conflict, we suggest that different forms of conflict (usually instigated by different types of conflict actors) are linked to seasonal changes in agricultural output through different pathways. Changes in political violence, especially that targeted against civilians, can be better understood through the *rapacity channel* wherein organized armed groups target civilians when and where it matters the most. Changes in social unrest can be best explained by the *opportunity cost channel* wherein farmers are less inclined to involve in protests and demonstrations when they are busy harvesting or have just observed harvest-related positive income. We also suppose the presence of the *resentment channel* that has the potential to reinforce the rapacity channel (a short-term increase in income inequality may trigger communal conflict) but mitigate the opportunity cost channel (non-farmers in predominantly agrarian regions may be provoked by the positive changes in income observed by farmers).

We estimate up to a nine percent increase in violence against civilians and up to a six percent increase in battles during harvest months relative to the rest of the year. These estimated effects are statistically significantly different from zero. We also estimate up to a thirteen percent decrease in protests and riots. While these estimated effects are of sizeable magnitude, they are statistically indistinguishable from zero. All these effects are amplified or more precisely estimated in rural cells, as well as in cells with rainfed agriculture.

In examining the mechanisms, we find that the increase in political violence, specifically violence against civilians, manifests after relatively good crop years, while a decrease in protests is evident after relatively bad crop years. This is a reassuring finding, particularly vis-à-vis our proposed effects governing different forms of conflict. An increase in political violence after a reasonably good harvest echoes the literature on food abundance and conflict (Koren, 2018; McGuirk and Burke, 2020). A decrease in protests only after a poor harvest is due to the opportunity cost channel not being offset by the resentment channel, as there is little change in harvest-related within-cell income inequality (between farmers and non-farmers) in those years.

This study offers insights into conflict resolution and development policies. First, both theoretically and empirically we make a case for one size does not fit all argument in conflict resolution: the harvest season is associated with opposite outcomes for different groups and different types of conflict. Organizations concerned about conflict writ large need to tailor their programs to the region, the type of conflict anticipated, and the groups involved in conflict. Recognizing that changes in political violence and social unrest in rural Southeast Asia are associated with rice harvest months can greatly assist in enhancing planning efforts by local governments and international agencies focused on rural development programs.

While much of the literature and indeed much development work are focused on the effects of food insecurity, our findings suggest that relative food *abundance* can be associated with an increase in political violence, and that food insecurity should not be the sole focus of conflict resolution, humanitarian, and development efforts. To the extent that this violence is driven by rapacious attacks on civilians—particularly farmers in crop-producing areas—those looking to protect civilians from violence can target their efforts more effectively.

The regularity of this finding—essentially that an increase in political violence may occur every year around the same time—can guide the timing and focus of development aid and external assistance to minimize conflict between groups. Good crop-producing years should be seen as times of special importance, given the increased chances of conflict. An expectation of seasonal conflict would lead to more robust preparations by stakeholders, both in terms of addressing the larger conflicts that can produce political violence, and in protecting farmers and agricultural surplus.

# APPENDICES

## APPENDIX B. TABLES

**Table A1: Harvest-time conflict in the croplands of Southeast Asia (subset of years)**

|  | All events | Battles | Violence | Riots | Protests |
|---|---|---|---|---|---|
| *Balanced panel: all countries, years 2018-2023* | | | | | |
| Cropland×Harvest | -0.120 | 0.100*** | 0.026 | -0.018** | -0.228* |
|  | (0.117) | (0.037) | (0.016) | (0.008) | (0.119) |
| Obs. | 27,072 | 27,072 | 27,072 | 27,072 | 27,072 |
| $R^2$ | 0.500 | 0.470 | 0.505 | 0.122 | 0.385 |
| *The magnitude of the estimated effect relative to the baseline conflict in the average cropland (%)* | | | | | |
| *Baseline conflict* | *2.42* | *0.96* | *0.43* | *0.05* | *0.98* |
| *Average cropland* | *0.78* | *0.78* | *0.78* | *0.78* | *0.78* |
| Cropland×Harvest | -3.9 | 8.1*** | 4.7 | -26.5** | -18.1* |
|  | (3.8) | (3.0) | (3.0) | (12.2) | (9.5) |

Note: The outcome variable is the number of conflict incidents in a cell in a year-month, and the treatment variable is the cropland size interacted with the indicator for the harvest season. The "All events" column combines all forms of conflict, and the "Battles" column battles and explosions/remote violence. The remaining columns represent the separate event types as labeled. All regressions include cell, country-year, and year-month fixed effects and control for contemporaneous rainfall. The values in parentheses are standard errors adjusted to clustering at the cell level; ***, **, and * denote 0.01, 0.05, and 0.10 statistical significance levels. The magnitudes of the effect, presented in percentage terms, are calculated as $\hat{\beta} \times \bar{a}/\bar{c} \times 100\%$, where $\hat{\beta}$ is the parameter estimate, $\bar{a}$ is the average size of the cropland, and $\bar{c}$ is the baseline conflict, which is, in effect, the unconditional mean of the outcome variable.

**Table A2: Harvest-time conflict in the croplands of Southeast Asia (subset of countries)**

|  | All events | Battles | Violence | Riots | Protests |
|---|---|---|---|---|---|
| *Balanced panel: All countries except Indonesia, Malaysia, and the Philippines, all years* | | | | | |
| Cropland×Harvest | -0.028 | 0.022 | 0.020*** | -0.003 | -0.067 |
|  | (0.056) | (0.018) | (0.007) | (0.004) | (0.058) |
| Obs. | 28,056 | 28,056 | 28,056 | 28,056 | 28,056 |
| $R^2$ | 0.375 | 0.376 | 0.334 | 0.074 | 0.273 |
| *The magnitude of the estimated effect relative to the baseline conflict in the average cropland (%)* | | | | | |
| Baseline conflict | 1.99 | 1.04 | 0.30 | 0.03 | 0.62 |
| Average cropland | 1.34 | 1.34 | 1.34 | 1.34 | 1.34 |
| Cropland×Harvest | -1.9 | 2.9 | 9.0*** | -15.9 | -14.7 |
|  | (3.8) | (2.3) | (3.0) | (19.4) | (12.6) |

Note: The outcome variable is the number of conflict incidents in a cell in a year-month, and the treatment variable is the cropland size interacted with the indicator for the harvest season. The "All events" column combines all forms of conflict, and the "Battles" column battles and explosions/remote violence. The remaining columns represent the separate event types as labeled. All regressions include cell, country-year, and year-month fixed effects and control for contemporaneous rainfall. The values in parentheses are standard errors adjusted to clustering at the cell level; ***, **, and * denote 0.01, 0.05, and 0.10 statistical significance levels. The magnitudes of the effect, presented in percentage terms, are calculated as $\hat{\beta} \times \bar{a}/\bar{c} \times 100\%$, where $\hat{\beta}$ is the parameter estimate, $\bar{a}$ is the average cropland size, and $\bar{c}$ is the baseline conflict in the croplands.

**Table A3: Harvest-time conflict in the croplands of Southeast Asia (0.5-degree cells)**

|  | All events | Battles | Violence | Riots | Protests |
|---|---|---|---|---|---|
| Cropland×Harvest | 0.010 | 0.060*** | 0.026*** | -0.005 | -0.070** |
|  | (0.040) | (0.016) | (0.007) | (0.003) | (0.036) |
| Obs. | 190,956 | 190,956 | 190,956 | 190,956 | 190,956 |
| $R^2$ | 0.296 | 0.211 | 0.284 | 0.085 | 0.265 |
| *The magnitude of the estimated effect relative to the baseline conflict in the average cropland (%)* | | | | | |
| *Baseline conflict* | *0.45* | *0.18* | *0.09* | *0.01* | *0.17* |
| *Average cropland* | *0.24* | *0.24* | *0.24* | *0.24* | *0.24* |
| Cropland×Harvest | 0.6 | 8.2*** | 6.8*** | -11.5 | -10.1** |
|  | (2.1) | (2.2) | (1.7) | (7.3) | (5.1) |

Note: The outcome variable is the number of conflict incidents in a cell in a year-month, and the treatment variable is the cropland area (10,000 ha) interacted with the indicator for the harvest season. The "All events" column combines all forms of conflict, and the "Battles" column battles and explosions/remote violence. The remaining columns represent the separate event types as labeled. All regressions include cell, country-year, and year-month fixed effects and control for contemporaneous rainfall. The values in parentheses are standard errors adjusted to clustering at the cell level; ***, **, and * denote 0.01, 0.05, and 0.10 statistical significance levels. The magnitudes of the effect, presented in percentage terms, are calculated as $\hat{\beta} \times \bar{a}/\bar{c} \times 100\%$, where $\hat{\beta}$ is the parameter estimate, $\bar{a}$ is the average cropland size, and $\bar{c}$ is the baseline conflict.

**Table A4: Harvest-time conflict in the croplands of Southeast Asia (rainfed vs irrigated)**

|                              | All events | Battles    | Violence  | Riots   | Protests  |
|------------------------------|------------|------------|-----------|---------|-----------|
| Cropland×Harvest             | -0.034     | 0.127***   | 0.052***  | -0.010  | -0.202**  |
|                              | (0.105)    | (0.040)    | (0.019)   | (0.008) | (0.097)   |
| Cropland×Harvest×Irrigated   | 0.037      | -0.191***  | -0.035    | 0.010   | 0.254**   |
|                              | (0.145)    | (0.056)    | (0.024)   | (0.010) | (0.127)   |
| Obs.                         | 49,236     | 49,236     | 49,236    | 49,236  | 49,236    |
| $R^2$                        | 0.388      | 0.373      | 0.354     | 0.113   | 0.303     |

Note: The outcome variable is the number of conflict incidents in a cell in a year-month, and the treatment variable is the cropland area (10,000 ha) interacted with the indicator for the harvest season. This treatment variable is interacted with the cell-specific proportion of irrigated cropland. The "All events" column combines all forms of conflict, and the "Battles" column battles and explosions/remote violence. The remaining columns represent the separate event types as labeled. All regressions include cell, country-year, and year-month fixed effects and control for contemporaneous rainfall. The values in parentheses are standard errors adjusted to clustering at the cell level; ***, **, and * denote 0.01, 0.05, and 0.10 statistical significance levels.

**Table A5: Harvest-time conflict in the croplands of Southeast Asia (rural vs urban)**

|  | All events | Battles | Violence | Riots | Protests |
|---|---|---|---|---|---|
| Cropland×Harvest×Urban | -0.009 | 0.025 | 0.016 | -0.004 | -0.045 |
|  | (0.056) | (0.024) | (0.016) | (0.004) | (0.056) |
| Cropland×Harvest×Rural | -0.022 | 0.052** | 0.047*** | -0.007 | -0.115 |
|  | (0.074) | (0.024) | (0.017) | (0.006) | (0.071) |
| Obs. | 49,236 | 49,236 | 49,236 | 49,236 | 49,236 |
| $R^2$ | 0.389 | 0.373 | 0.354 | 0.113 | 0.302 |

Note: The outcome variable is the number of conflict incidents in a cell in a year-month, and the treatment variable is the cropland area (10,000 ha) interacted with the indicator for the harvest season. This treatment variable is further interacted with the indicator variable for rural and urban cells. A cell is deemed *urban* if it is located within one hour of the nearest city center; it is deemed *rural*, otherwise. The "All events" column combines all forms of conflict, and the "Battles" column battles and explosions/remote violence. The remaining columns represent the separate event types as labeled. All regressions include cell, country-year, and year-month fixed effects and control for contemporaneous rainfall. The values in parentheses are standard errors adjusted to clustering at the cell level; ***, **, and * denote 0.01, 0.05, and 0.10 statistical significance levels.

**Table A6: The effect of growing season rainfall on the harvest-time conflict**

|  | All events | Battles | Violence | Riots | Protests |
|---|---|---|---|---|---|
| Cropland×Harvest | 0.057 | 0.387** | -0.027 | -0.015 | -0.288* |
|  | (0.255) | (0.164) | (0.070) | (0.011) | (0.162) |
| Cropland×Harvest×Rain | -0.031 | -0.661** | 0.135 | 0.016 | 0.479** |
|  | (0.423) | (0.318) | (0.142) | (0.015) | (0.216) |
| Cropland×Harvest×Rain$^2$ | 0.071 | 0.244* | -0.062 | -0.005 | -0.248*** |
|  | (0.173) | (0.137) | (0.059) | (0.007) | (0.089) |
| Obs. | 49,236 | 49,236 | 49,236 | 49,236 | 49,236 |
| R$^2$ | 0.299 | 0.249 | 0.308 | 0.106 | 0.262 |

Note: The outcome variable is the number of conflict incidents in a cell in a year-month, and the treatment variable is the cropland area (10,000 ha) interacted with the indicator for the harvest season. This treatment variable is further interacted with the linear and quadratic terms of the monthly average growing season rainfall (100mm). The "All events" column combines all forms of conflict, and the "Battles" column battles and explosions/remote violence. The remaining columns represent the separate event types as labeled. All regressions include cell, country-year, and year-month fixed effects and control for contemporaneous rainfall. The values in parentheses are standard errors adjusted to clustering at the cell level; ***, **, and * denote 0.01, 0.05, and 0.10 statistical significance levels.

**Table A7: The effect of growing season conflict on the harvest-time conflict**

|  | All events | Battles | Violence | Riots | Protests |
|---|---|---|---|---|---|
| Conflict | -0.104 | 1.145*** | 0.300*** | 0.171*** | 0.199** |
|  | (0.225) | (0.301) | (0.074) | (0.017) | (0.087) |
| Cropland×Harvest | 0.007 | 0.010 | 0.022 | -0.004 | -0.055 |
|  | (0.046) | (0.021) | (0.015) | (0.004) | (0.041) |
| Cropland×Harvest×Conflict | -0.046 | 0.300 | 0.085* | -0.018 | -0.074 |
|  | (0.060) | (0.226) | (0.050) | (0.034) | (0.046) |
| Obs. | 49,236 | 49,236 | 49,236 | 49,236 | 49,236 |
| R² | 0.389 | 0.379 | 0.355 | 0.124 | 0.303 |

Note: The outcome variable is the number of conflict incidents in a cell in a year-month, and the treatment variable is the cropland area (10,000 ha) interacted with the indicator for the harvest season. This treatment variable is further interacted with the indicator variable depicting the incidence of any form of conflict during the crop growing season. The "All events" column combines all forms of conflict, and the "Battles" column battles and explosions/remote violence. The remaining columns represent the separate event types as labeled. All regressions include cell, country-year, and year-month fixed effects and control for contemporaneous rainfall. Values in parentheses are standard errors adjusted for clustering at the cell level; ***, **, and * denote 0.01, 0.05, and 0.10, respectively. [a] denotes values smaller than the entries.

APPENDIX B. FIGURES

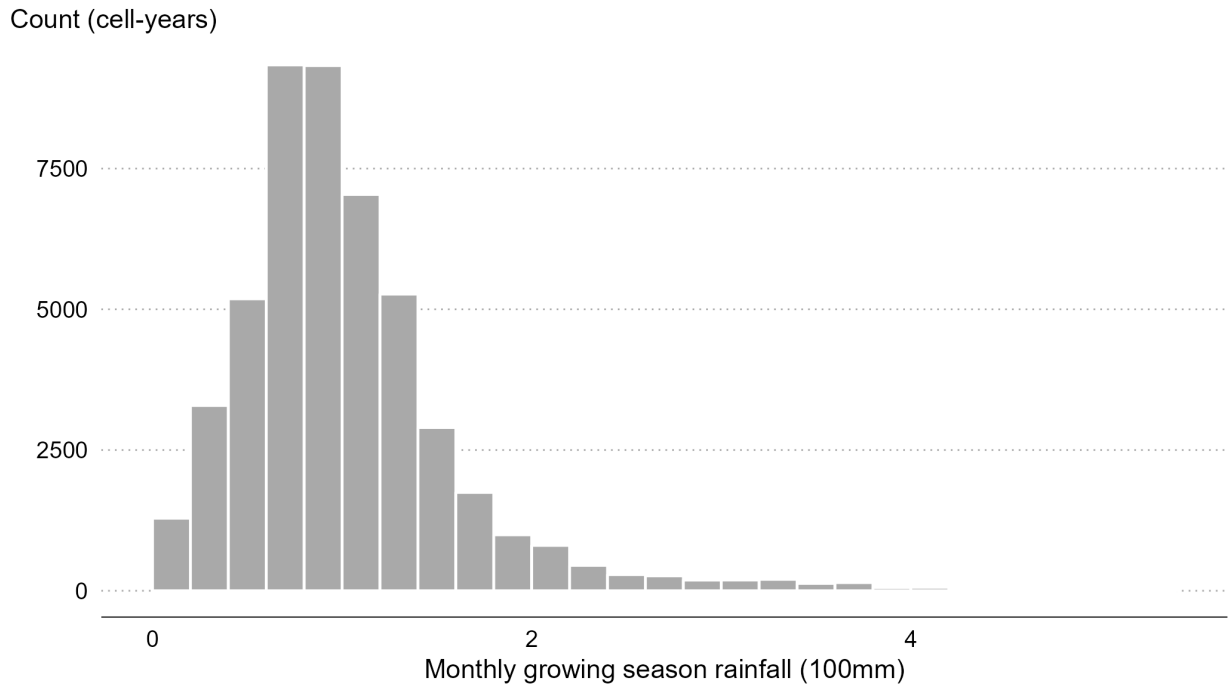

**Figure B1: Distribution of the monthly average growing season rainfall**
Note: Data are from the CPC Global Unified Gauge-Based Analysis of Daily Precipitation data provided by the National Oceanic and Atmospheric Administration's Physical Sciences Laboratory (https://psl.noaa.gov) and Sacks et al. (2010) and cover the years 2010 through 2023 (unless otherwise indicated) from Cambodia, Indonesia (from 2015), Laos, Malaysia (from 2018), Myanmar, the Philippines (from 2016), Thailand, and Vietnam.

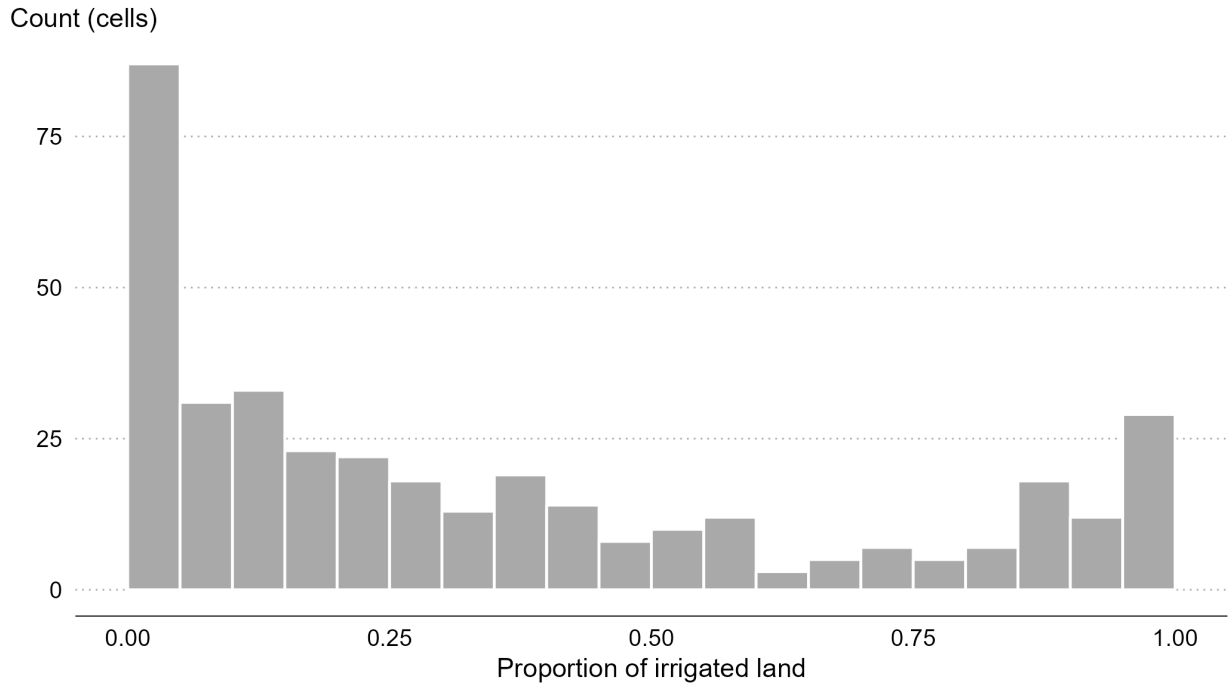

**Figure B2: Distribution of the irrigated land**
Note: Data are from IFRPI (2019) and cover Cambodia, Indonesia, Laos, Malaysia, Myanmar, the Philippines, Thailand, and Vietnam.

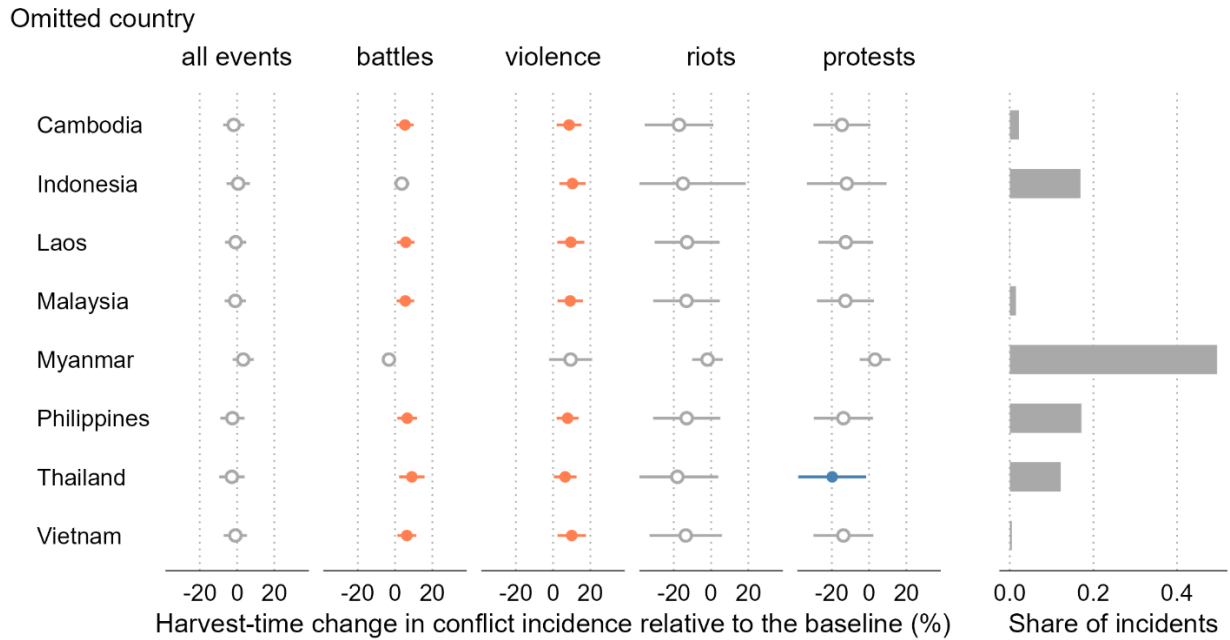

**Figure B3: Sensitivity of the estimates to omitting one country at a time from the dataset**

Note: The dots indicate point estimates, and the error bars show 95% confidence intervals around the point estimates of the effect calculated as $\hat{\beta} \times \bar{a}/\bar{c} \times 100\%$, where $\hat{\beta}$ is the parameter estimate from Equation (1), $\bar{a}$ is the average size of the cropland, and $\bar{c}$ is the average number of conflict incidents. Confidence intervals are obtained using standard errors adjusted to clustering at the cell level. The colored dots and error bars show statistically significant positive (orange) or negative (blue) impacts at the 5% level. The horizontal bars (far-right panel) depict the relative prevalence of conflict (all events) by country.

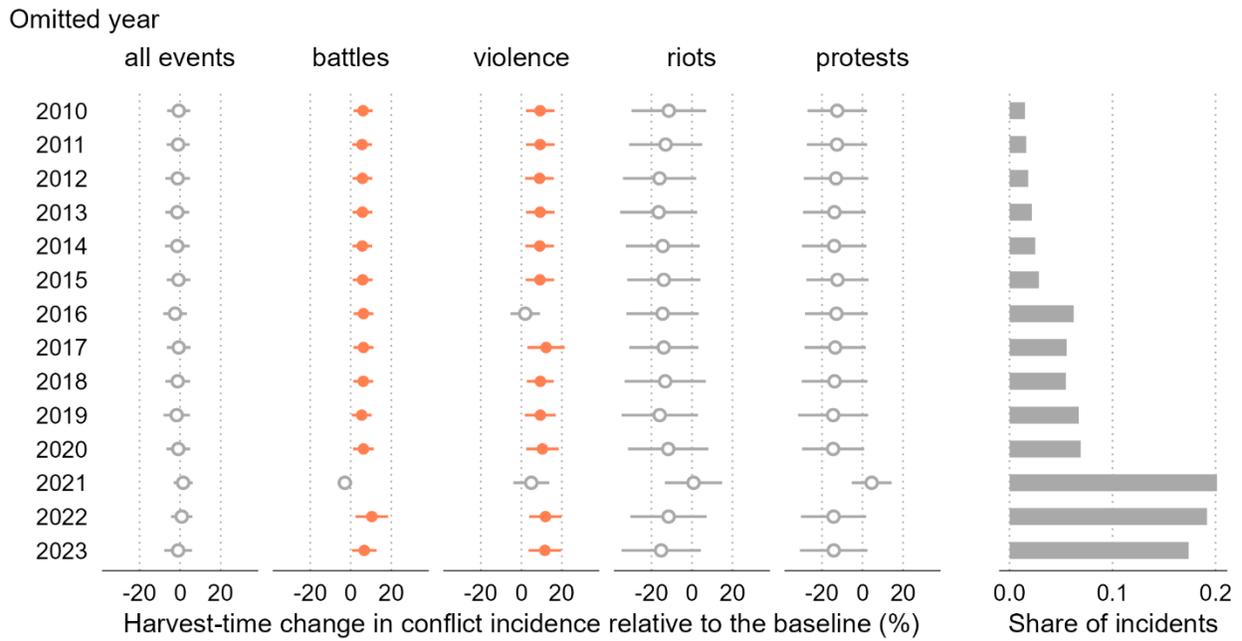

**Figure B4: Sensitivity of the estimates to omitting one year at a time from the dataset**

Note: The dots indicate point estimates, and the error bars show 95% confidence intervals around the point estimates of the effect calculated as $\hat{\beta} \times \bar{a}/\bar{c} \times 100\%$, where $\hat{\beta}$ is the parameter estimate from Equation (1), $\bar{a}$ is the average size of the cropland, and $\bar{c}$ is the average number of conflict incidents. The confidence intervals are obtained using standard errors adjusted to clustering at the cell level. The colored dots and error bars show statistically significant positive (orange) or negative (blue) impacts at the 5% level. The horizontal bars (far-right panel) depict the relative prevalence of conflict (all events) by year.

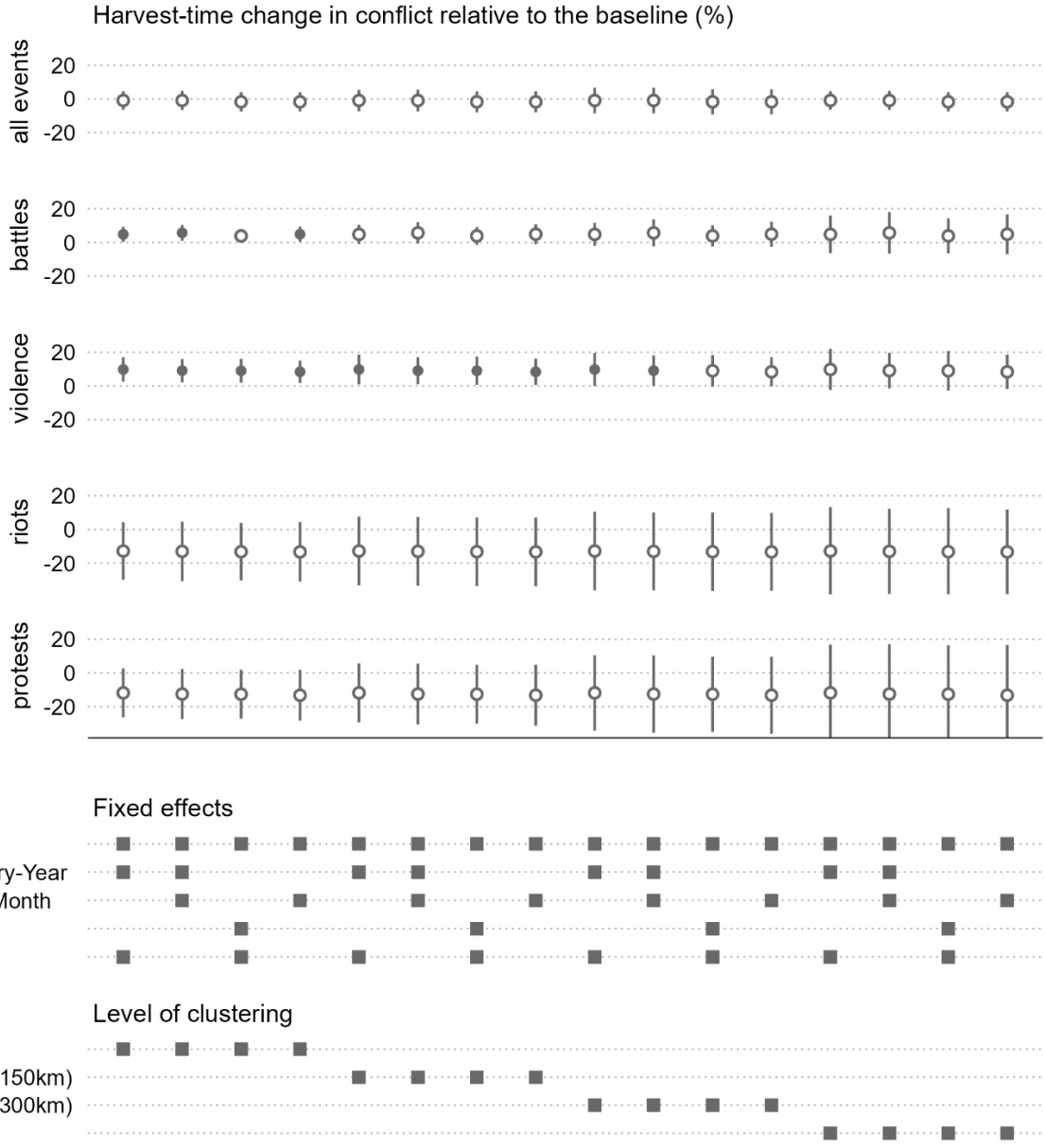

**Figure B5: Specification chart**

Note: In the top panel, the dots are point estimates, and the error bars show 95% confidence intervals around the point estimates of the effects calculated as $\hat{\beta} \times \bar{a}/\bar{c} \times 100\%$, where $\hat{\beta}$ is the parameter estimate from Equation (1), $\bar{a}$ is the average size of the cropland, and $\bar{c}$ is the average number of conflict incidents. The filled dots indicate the effects that are statistically significantly different from zero at the 5% level. The bottom panel shows different combinations of fixed effects and levels of error clustering.

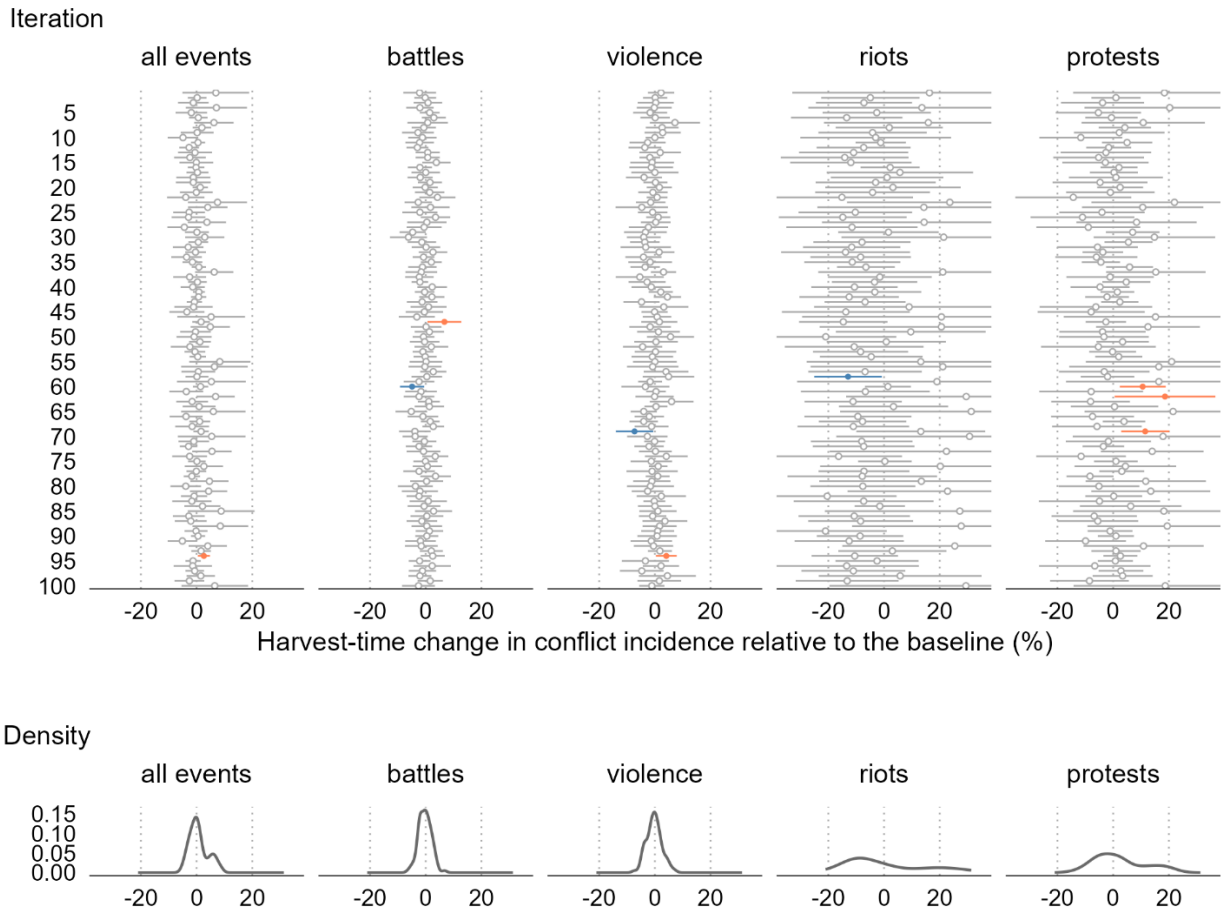

**Figure B6: Estimated impacts using randomly assigned harvest seasons**

Note: In the top panel, the dots are point estimates, and the error bars show 95% confidence intervals around the point estimates of the effects calculated as $\hat{\beta} \times \bar{a}/\bar{c} \times 100\%$, where $\hat{\beta}$ is the parameter estimate from Equation (1), $\bar{a}$ is the average size of the cropland, and $\bar{c}$ is the average number of conflict incidents. The confidence intervals are obtained using standard errors adjusted to clustering at the cell level. The colored dots and error bars show the impacts that are statistically significantly different from zero at the 5% level. The bottom panel shows the densities of the point estimates.